\documentclass[structabstract]{aa}  
\usepackage{graphicx}
\usepackage{txfonts}
\begin{document}
   \title{The circumstellar disk of HH 30}

   \subtitle{Searching for signs of disk evolution with multi-wavelength modeling}

   \author{D. Madlener\inst{1},
           S. Wolf\inst{1},
           A. Dutrey\inst{2,3},
           S. Guilloteau\inst{2,3}
          }

   \institute{Christian-Albrechts-Universit\"at zu Kiel,
              Institut f\"ur Theoretische Physik und Astrophysik (ITAP),
              Leibnizstr. 15,
              24118 Kiel,
              Germany\\
              \email{[dmadlener;wolf]@astrophysik.uni-kiel.de}
            \and 
              Universit\'e de Bordeaux, 
              Observatoire Aquitain des Sciences de l'Univers, 
              2 rue de l'Observatoire, 
              BP 89, F-33271 Floirac Cedex, France                
            \and               
              CNRS,
              UMR 5804, 
              Laboratoire d'Astrophysique de Bordeaux, 
              2 rue de l'Observatoire, 
              BP 89, F-33271 Floirac Cedex, France              
              \email{[Anne.Dutrey;Stephane.Guilloteau]@obs.u-bordeaux1.fr}
             }

   \date{}

 
  \abstract
   {Circumstellar disks are characteristic for star formation and vanish during the first few Myr of stellar evolution. During this time planets are believed to form in the dense midplane by growth, sedimentation and aggregation of dust. Indicators of disk evolution, such as holes and gaps, can be traced in the spectral energy distribution (SED) and spatially resolved images.}
   {We aim to construct a self-consistent model of HH 30 by fitting all available continuum observations simultaneously. New data sets not available in previous studies, such as high-resolution interferometric imaging with the Plateau de Bure Interferometer (PdBI) at $\lambda{=}1.3$mm and SED measured with IRS on the \emph{Spitzer Space Telescope} in the mid-infrared, put strong constraints on predictions and are likely to provide new insights into the evolutionary state of this object. }
   {A parameter study based on simulated annealing was performed to find unbiased best-fit models for independent observations made in the wavelength domain $\lambda\sim1\mu\textrm{m}\ldots4$mm. The method essentially creates a Markov chain through parameter space by comparing predictions generated by our self-consistent continuum radiation transfer code \texttt{MC3D} with observations.} 
   {We present models of the edge-on circumstellar disk of HH 30 based on observations from the near-infrared to mm-wavelengths that suggest the presence of an inner depletion zone with ${\sim}45$AU radius and a steep decline of mm opacity beyond ${\gtrsim}140$AU. Our modeling indicates that several modes of dust evolution such as growth, settling, and radial migration are taking place in this object.}
   {High-resolution observations of HH 30 at different wavelengths with next-generation observatories such as ALMA and JWST will enable the modeling of inhomogeneous dust properties and significantly expand our understanding of circumstellar disk evolution.}


   \maketitle
%

\section{Introduction}

Several theories have been proposed during the last decades to explain the formation of planets in protostellar disks, and most hypotheses are based on growth and sedimentation of dust grains, e.g. the core accretion scenario (Papaloizou \& Terquem 2006; Lissauer \& Stevenson 2007). Nanometer-sized dust grains grow through low-velocity collisions by several orders of magnitude (Beckwith et al. 2000; Dominik et al. 2007; Natta et al. 2007). During this coagulation process bigger dust grains ($\gg 1\mu$m) decouple from the turbulent gas flow and settle in the mid-plane of the disk (Fromang \& Papaloizou 2006). This sedimentation process is expected to form a thin embedded disk consisting of larger aggregates (Dubrulle et al. 1995; Carballido et al. 2006). It is still debated how nature circumvents the 'meter size barrier' that should effectively remove boulders with diameters $d\sim$1m from this inner disk by gas drag and accretion onto the protostar (Weidenschilling 1977; Brauer et al. 2008) or fragmentation due to high-speed collisions (Blum et al. 2000). Direct observations of these boulders are impossible with current instruments but indirect signs of coagulation and stratification are available. The spectral energy distribution (SED) provides constraints on dust mass, grain size distribution and total mass. High-resolution images in scattered light from optical to mid-infrared (MIR) reveal attributes of small dust grains through wavelength-dependent opacity, and interferometric continuum observations in the (sub)mm regime are used to model the spatial dust distribution of larger grains (D'Alessio et al. 2001; Dullemond \& Dominik 2004; Watson et al. 2007). Resolved images at different wavelengths and good coverage of the SED are mandatory to reduce model degeneracies (Chiang et al. 2001).

HH 30 is a young stellar object (YSO) located in the dark cloud L1551 at a distance of ${\sim}140$pc in Taurus. The accreting protostar features an edge-on indirectly illuminated flared disk. In the optical and near-infrared (NIR) this disk appears as a nebulosity separated by an obscuring belt with wavelength-dependent thickness (Burrows et al. 1996, hereafter B96; Cotera et al. 2001, hereafter C01). Interferometric observations of $^{13}$CO (J = 2-1) are consistent with a gaseous disk in Keplerian rotation around an enclosed mass of $(0.45{\pm}0.04)$M$_\odot$ that corresponds to a typical TTauri star with spectral class M0${\pm}1$ (Pety et al. 2006, hereafter P06). Observed photometric and polarimetric variability of HH 30 shows periodicity on the scale of a few days and seems to be caused by either a light-house effect, periodic shadowing by infalling matter, or a close companion (Watson \& Stapelfeldt 2007; Duran-Rojas et al. 2009). The impressive dynamic bipolar jet features an undulating morphology and ballistic motion that can be explained by orbital motion or precession of the source due to gravitational interaction with a partner separated by ${<}18$AU (Anglada et al. 2007). The jet is associated with an entrained conical molecular outflow featuring an opening angle of $\sim$30$^\circ$ (P06), whose morphology is consistent with a close binary (Tambovtseva et al. 2008). Interferometric imaging in the continuum at $\lambda=1.3$mm resolved a region of reduced brightness at the center of the system, suggesting a hole in the dust distribution of $\sim$40AU radius. This inner zone may have been depleted of dust by tidal clearing of an unresolved binary with a separation of ${\sim}15$AU and comparable mass, e.g. $M_1=0.3$M$_\odot$ and $M_2=0.15$M$_\odot$ (Guilloteau et al. 2008, hereafter G08). In this case, the binary components would be cooler ($T_{\rm{eff}}{\sim}3500$K) and younger ($t{\sim}$1Myr) with a spectral type of ${\sim}$M3. 

The aim of our campaign is to construct a model of HH 30 consistent with all available observations from the NIR to mm-wavelengths. As pointed out earlier, only simultaneous optimization of independent observations, i.e. the well-covered SED, spatially resolved images and interferometric visibilities, can significantly reduce model ambiguities. An analysis of the results will enable us to confirm signs of disk evolution, e.g. the presence of an inner hole, dust growth, and settling. We will also be able to predict possible insights with the upcoming next generation of observatories.


\section{Observations}
\label{observations}
\begin{figure}[tb!]
  \centering
  \includegraphics[width=0.5\textwidth]{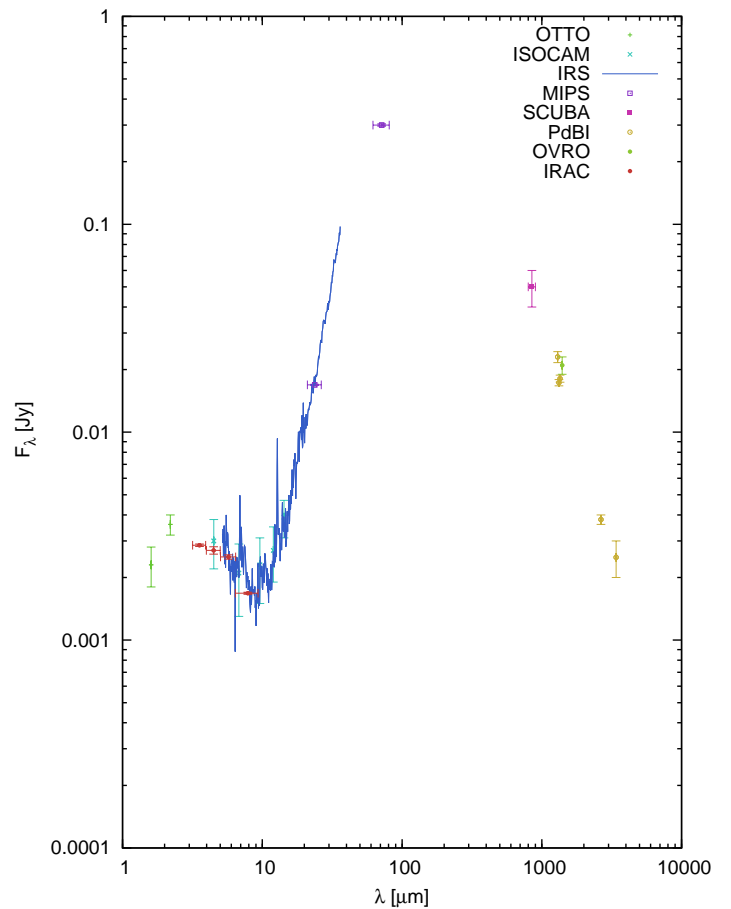}
  \caption{SED of HH 30. The data were taken from the following sources: Otto bolometer on KPNO: Vrba et al. (1985), ISOCAM/ISO: Brandner et al. (2000), IRS/SST: Furlan et al. (2008), MIPS,IRAC/SST: \emph{Spitzer Heritage Archive}, see Table \ref{tab:Spitzer_photometry}, SCUBA/JCMT: Moriarty-Schieven et al. (2006), PdBI: Pety et al. (2006), OVRO: Padgett et al. (2001).}
  \label{hh30_sed}
\end{figure}

HH 30 has been studied extensively over the last decades from the optical band to mm-wavelengths. Most research has been conducted in the optical and NIR band including photometry, polarimetry, medium-resolution imaging and high-resolution spectroscopy with ground-based instruments (Vrba et al. 1985; Appenzeller et al. 2005; Anglada et al. 2007) and high-resolution imaging with the \emph{Hubble Space Telescope} (HST) from outside the atmosphere (B96; Stapelfeldt et al. 1999; C01; Watson \& Stapelfeldt 2004, hereafter W04). The SED in the MIR has been measured with ISOCAM/ISO (Stapelfeldt \& Moneti 1999; Brandner et al. 2000) and refined substantially by observations with IRS on the \emph{Spitzer Space Telescope} (SST, Furlan et al. 2008). Mapping of continuum fluxes at (sub)mm-wavelengths, observations of rotational transitions of CO isotopes and HCO$^+$, and high-resolution imaging with interferometers have provided valuable insights revealing the dynamics of the gaseous component, its temperature distribution and the very low spectral index $\beta_{\rm{mm}}{\sim}0.4$ of the dust (Reipurth et al. 1993; Padgett et al. 2001; Moriarty-Schieven et al. 2006; P06; G08). We will present the most prominent observations used for this study in the following paragraphs.

\subsection{IRAM Plateau de Bure Interferometer}
\label{pdbi_imaging}

\begin{figure}
  \centering
  \includegraphics[width=0.5\textwidth]{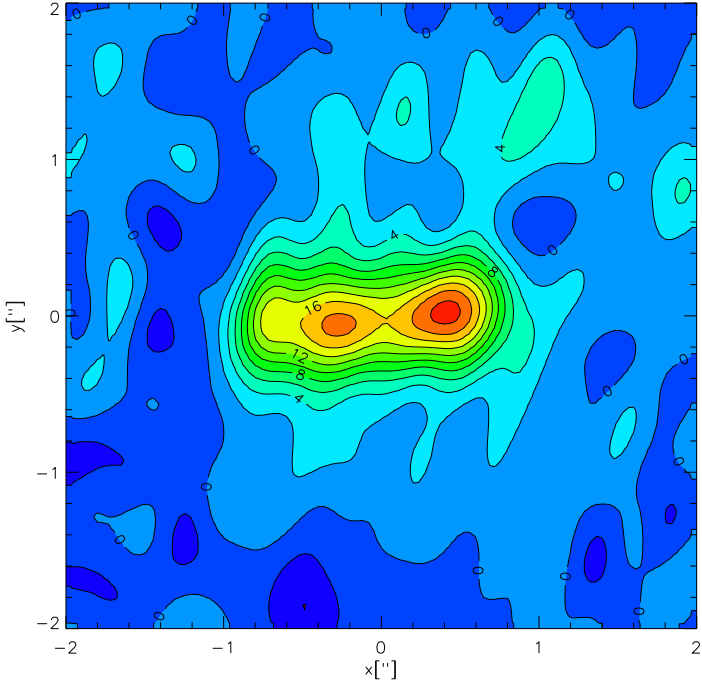}
  \caption{Interferometric image of HH 30 at $\lambda$=1.3mm published in G08. The center lies at $\alpha = 04^h31^m37^s.469$ and $\delta = 18^\circ 12'24''.22$ in J2000. Every contour line indicates a brightness step of 2$\sigma{\sim}0.46$mK with the right-hand side maximum peaking at $0.43$K or ${\sim}19\sigma$, the two peaks are separated by ${\sim}100$AU.}
 \label{hh30_sigma}
\end{figure}

High angular resolution observations of HH 30 at $\lambda = 1.3$mm were conducted on 2007 February 3 with the IRAM Plateau de Bure Interferometer (PdBI) using six antennas in the A configuration for a total on-source observing time of 6h using dual polarization and a total effective bandwidth of ${\sim}1.8$GHz. Observations of the calibrators indicated a seeing better than ${\sim}0.15"$ after calibration. The rms phase noise stayed between $19^\circ$ and $53^\circ$, the brightness rms noise was 0.185mJy/beam or 23mK with a peak brightness of 0.43K corresponding to ${\sim}19\sigma$. Data reduction was performed with \texttt{GILDAS}, a radioastronomical software package developed at IRAM (Pety 2005). Natural weighting was used to produce the image presented in Fig.~\ref{hh30_sigma}. The beam has a FWHM extension of $0.59"{\times}0.32"$ at PA = $22^\circ$, yielding a linear resolution along the mid-plane of ${\sim}45$AU and ${\sim}83$AU in the perpendicular direction at HH 30's distance of ${\sim}140$pc. Additional details on this observation can be found in G08.

\subsection{IRS/SST}

HH 30 was observed with the IRS spectrograph on board the SST on 2004 February 8 using the modules Short-Low [SL] from 5.2-14$\mu$m and Long-Low [LL] from 14-38$\mu$m with spectral resolution $\lambda/\Delta\lambda{\sim}90$ using the stare mode (AOR key: 3552512, PI: J.~R. Houck). The data was reduced with SMART (Higdon et al. 2004), which interpolates bad pixels and subtracts the sky (Furlan et al. 2008). The SL2 part of the spectrum was too weak for direct extraction and was not published, but re-reduction of the raw data using an off-order sky subtraction allowed subsequent extraction and calibration of the SL2 spectrum from 7.35$\mu$m to 8.24$\mu$m (private communication with E. Furlan).

\subsection{IRAC/SST}

The dark cloud L1551 was imaged on 2004 October 7 with the IRAC instrument using the map mode in all four available channels at 3.5$\mu$m, 4.5$\mu$m, 5.8$\mu$m and 8.0$\mu$m (AOR key: 3653120, PI: G. Fazio). We acquired the BCD sets from the \emph{Spitzer Heritage Archive} (SHA) v2.0 and extracted the fluxes with MOPEX v18.4.9 using the Mosaic and APEX/single-source pipeline. HH 30 was detected in all four maps as an extended source, the fluxes determined by aperture photometry are listed in Table \ref{tab:Spitzer_photometry}.

\subsection{MIPS/SST}

The surroundings of HH 30 were revisited on 2006 February 16 using the MIPS instrument in scan mode with medium rate in all three available channels at 24$\mu$m, 70$\mu$m and 160$\mu$m (AOR key: 12615168, PI: G. Fazio). Congruent with the analysis of the IRAC data we acquired the BCD sets from the SHA and extracted the fluxes with MOPEX v18.4.9 using the Mosaic and APEX/single source pipeline. Our target appeared as a point source in the 24$\mu$m and 70$\mu$m maps, unfortunately it could not be distinguished from the background at 160$\mu$m. The resulting fluxes are listed in Table \ref{tab:Spitzer_photometry}.

\begin{table}
\caption{Photometric fluxes in the mid- and far-infrared observed with the IRAC and MIPS instruments on the SST.}
\centering
\begin{tabular}{ll}
 $\lambda$ [$\mu$m]& $F_\lambda$ [mJy] \\
\hline
\hline
$3.6 \pm 0.4$ & $2.86  \pm 0.04$ \\
$4.5 \pm 0.5$ & $2.70  \pm 0.11$ \\
$5.7 \pm 0.7$ & $2.51  \pm 0.06$ \\
$7.9 \pm 1.5$ & $1.68  \pm 0.02$ \\
$23.7\pm 2.7$ & $16.9  \pm 0.1$  \\
$71.4\pm 9.5$ & $300.2 \pm 3.9$ \\
\hline
\end{tabular}
\label{tab:Spitzer_photometry}
\end{table}

\subsection{NIC2/HST}

\begin{figure}[tb!]
  \centering
  \includegraphics[width=0.5\textwidth]{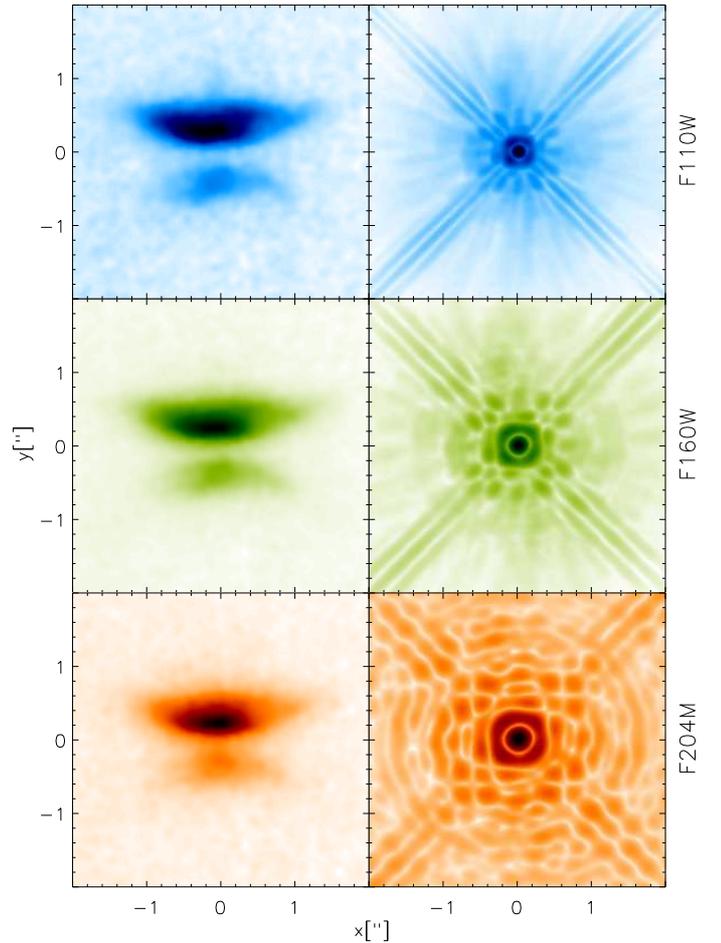}
  \caption{Inverse images made with NIC2 using filters F110W, F160W and F204M aligned with the corresponding point spread functions (PSF) synthesized with \texttt{Tiny Tim v7.0} (Krist 1995). The PSFs were scaled by the 5th root to enhance visual contrast, the images are unscaled and rotated by 32$^\circ$ to align the disk.}
  \label{fig:nic2_images}
\end{figure}

Our object was examined by the HST on 1997 September 29 with the NIC2 camera of the NICMOS instrument (Thompson et al. 1998). The observations were performed using fixed aperture and MULTIACCUM read-out mode with the filters F110W, F160W, F187N, F204M, and F212N (HST Proposal 7228, PI: E. Young). This dataset has already been employed in several efforts to model HH 30 (C01; Wood et al. 2002, hereafter W02; W04). For our study we used the three images obtained with the mid and wide filters F110W, F160W, and F204M. The three data sets are offered by the Hubble Legacy Archive (HLA) in its data release 4 (DR4), s. Fig.~\ref{fig:nic2_images}. This release applies the latest standard pipeline for level 2 products, consisting of calibration with Calnica 4.4.0 and distortion correction, cosmic ray removal and image combination with MultiDrizzle 3.1.0.

\section{Modeling}

This section presents all models, ingredients and tools necessary to follow our approach. At the studied scale, the dust dominates
the transport of radiation and thermal structure of the disk (Chiang \& Goldreich 1997). Using one chemical mixture and the simple approach of Mie scattering by spherical particles, our modeling effort concentrates on the spatial arrangement and grain size distribution of the dust. Approximating dust grains as perfect spheres made from one single compound may seem too optimistic because grains are expected to have a fractal structure and varying composition. Then again, as has been shown by Voshchinnikov (2002), size, shape, and chemical composition cannot be separated from one another, hence we restrict our modeling effort to the simplest shape available and one chemical recipe. Another necessary simplification is to neglect asymmetries and variations of light sources, because the available data are not tangible enough to support a sophisticated model that encompasses the jet, an extended protostar with accretion shocks, hot and cold spots, or even multiple stars without introducing ambiguities in abundance. With these limitations in mind we proceed with the description of our model.

\subsection{The disk}
\label{disk_model}

\begin{figure}
  \centering
  \includegraphics[width=0.5\textwidth]{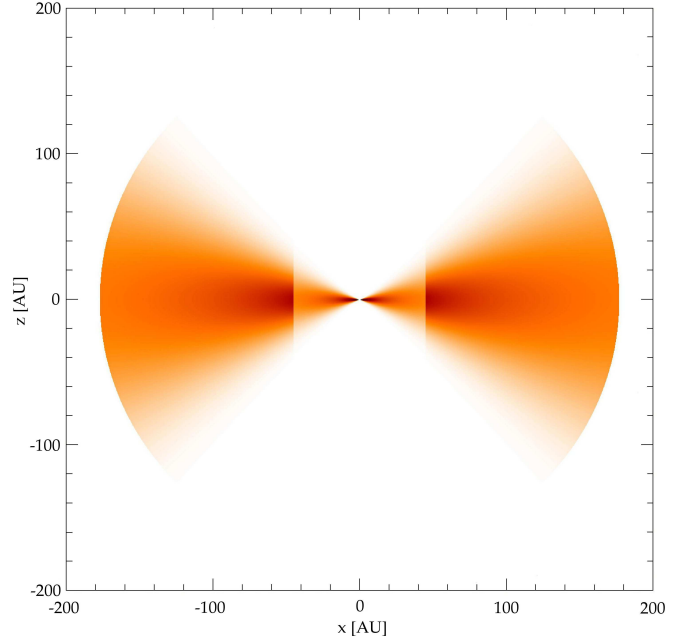}
  \caption{Cut through the $xz$-plane of the density distribution with depletion zone described by Eqn. (\ref{flared_disk}), (\ref{scale_height}) and (\ref{step_function}), effectively showing two interlaced disks. In our study the outer disk dominates the total mass and hence mm flux, while the inner disk contains warm dust necessary to reproduce the flux in the MIR. Parameters are taken from our best-fit model D, see Table \ref{tab:best_fit}.}
  \label{density_diagram}
\end{figure}

The standard model for an accreting disk is the $\alpha$-viscosity model introduced by Shakura \& Sunyaev (1973) and we use in this study a flared disk based on their ideas. The following parametrization assumes cylindrical symmetry: 

\begin{equation}
\rho_{\rm{dust}}(R,z) \sim {R}^{-\alpha} \exp\left[-\frac{z^2}{2h^2}\right].
\label{flared_disk}
\end{equation}
This ansatz allows flaring by varying the scaling height according to

\begin{equation}
h(R) = h_{\rm{100}} \left(\frac{R}{100\rm{AU}}\right)^\beta,
\label{scale_height}
\end{equation}
with the scaling height $h_{100}$ at a distance of 100AU and the flaring exponent $\beta$. This canonical model has been used succesfully in previous studies, e.g. the Butterfly Star (Wolf et al. 2003) and CB 26 (Sauter, Wolf et al. 2009), but observational constraints require a modification. The available mm observations of HH 30 can be explained by a hole of ${\sim}40$AU radius (G08) but the SED in the MIR suggests the presence of warm dust ($T{\sim}$200K). We hence modified (\ref{flared_disk}) with the step function

\begin{equation}
\theta(R) = \left\{
\begin{array}{rl}
\eta & : R_{\rm{in}}\leq R \leq R_{\rm{att}} \\
1 & : R_{\rm{att}}< R \leq R_{\rm{out}} \\
0 & : \textrm{ else, } \\
\end{array}
\right.
\label{step_function}
\end{equation}

thereby introducing an attenuated inner region spanning from $R_{\rm{in}}$ to $R_{\rm{att}}{<}R_{\rm{out}}$. The density jumps at this radius by the \emph{attenuation} $\eta\in[0,1]$, effectively separating the system in an inner and an outer disk, see Fig.~\ref{density_diagram}. The dip in mm brightness observed by G08 is compatible with two scenarios, either shadowing of the inner region by optically thick layers (e.g. Wolf et al. 2008) or depletion of dust by a dissipation process, e.g. photoevaporation, tidal interaction with a central binary, or planet formation. The low peak brightness of ${\sim}0.4$K in the mm map suggests the latter variant to be true (G08), see Fig.~\ref{hh30_sigma}. Nevertheless, this simple extension of the standard model is in principle compatible with both settings. The surface density

\begin{equation}
\Sigma_{\rm{dust}}(R) = \int_{-\infty}^{\infty} \rho_{\rm{dust}}(R,z) dz
\end{equation}
follows a power law $\Sigma_{\rm{dust}}\sim R^{-p}$ with $p = \alpha -\beta$. From accretion physics we expect 
\begin{equation}
\alpha = 3(\beta - \frac{1}{2}),
\label{eq:alpha_beta}
\end{equation}
which confines the radial and flaring exponents. However, we treated both parameters as independent quantities, because dust grains can decouple from the gas stream and are subjected to physical processes not included in the $\alpha$-disk model. Integrating (\ref{flared_disk}) over the entire model space yields 
\begin{equation}
\frac{m_{\rm{in}}}{m_{\rm{out}}} = \eta\cdot \left\{
\begin{array}{rl}
 \frac{R_{\rm{att}}^{1-p} - R_{\rm{in}}^{1-p}}{R_{\rm{out}}^{1-p} - R_{\rm{att}}^{1-p}} & \textrm{ for } p\neq 1 \\
  & \\
\frac{\ln[R_{\rm{att}}/R_{\rm{in}}]}{\ln[R_{\rm{out}}/R_{\rm{att}}]} & \textrm{ for } p = 1 \\
\end{array}
\right.
\end{equation}
for the mass ratio of inner and outer disk. The total dust mass is adjusted to any required value $m_{\rm{dust}}$ by multiplication of (\ref{flared_disk}) with an appropriate constant. 
Our model hence uses seven geometrical parameters $\eta$, $\alpha, \beta$, $R_{\rm{in}}$, $R_{\rm{att}}$, $R_{\rm{out}}$ and $h_{\rm{100}}$ to describe the shape and mass distribution of the disk and one additional parameter $m_{\rm{dust}}$ to fix the total dust mass. 

\subsection{Heating}

Three main sources of heating are known in protostellar systems, 
\begin{enumerate}
\item viscous dissipation in the disk, 
\item accretion onto the protostar, and 
\item contraction of the latter.
\end{enumerate}
We neglect heating of the disk by viscous dissipation because the energy production in the outer parts of a protostellar disk is about one order of magnitude lower than the star's contribution (Wolf et al. 2003). Because we are not interested in asymmetries, UV and X-ray excess, or line emission, we illuminated the system with a single black body positioned at the origin requiring two additional parameters, the effective temperature $T_{\rm{eff}}$ and bolometric luminosity $L_*$. Owing to obscuration by the disk, the illuminating source of the system cannot be observed directly, which leads to weak constraints on these parameters. 

\subsection{Dust}
It is sufficient to describe the mean grain shape, size distribution, and chemical composition because only mean optical properties of grain ensembles are used. The following simplifications were employed to reduce the number of free parameters necessary to describe a particular dust species:

\begin{enumerate}
\item spherically shaped grains,
\item homogeneous chemical composition, and
\item no spatial dependence of its properties.
\end{enumerate}
Given the index of refraction, the optical attributes of a homogeneous sphere can be calculated with Mie's solution to arbitrary precision. By averaging over a given grain size distribution and chemical composition, a single catalog can be deduced that summarizes all optical properties necessary to model the radiation transfer of a particular dust species. This numerical procedure partitions the interval $[a_{\rm{min}},a_{\rm{max}}]$  logarithmically and evaluates a weighted sum (Wolf 2003).

\paragraph{Size distribution}
\label{size_distr}
The standard ansatz for modeling a grain size distribution is a truncated power law 

\begin{equation}
n(a) \sim a^{-d}
\end{equation}
with the grain size exponent $d$ and the limiting grain sizes $a_{\rm{min}}$ and $a_{\rm{max}}$ (Mathis, Rumpl \& Nordsieck 1977, hereafter MRN). From these three parameters every average property of the ensemble can be deduced by integrating, e.g.

\begin{equation}
\left<a^n\right> = \frac{1-d}{a_{\rm{max}}^{1-d} - a_{\rm{min}}^{1-d}}\cdot\left\{
\begin{array}{rl}
\frac{a_{\rm{max}}^{n+1-d} - a_{\rm{min}}^{n+1-d}}{n+1-d} & :\,\, n\neq d-1 \\
 & \\
\ln\left[a_{\rm{max}}/a_{\rm{min}}\right] & :\,\, n = d-1. \\
\end{array}
\right.
\end{equation}
The standard grain size distribution introduced by MRN to explain the interstellar absorption has the parameter values $a_{\rm{min}} = 5$nm, $a_{\rm{max}}=250$nm, and $d=3.5$.

\paragraph{Chemical composition}


A homogeneous mixture of smoothed astrophysical silicate and graphite with a ratio of 5:3 for Si:C and a mean density of $\rho_{grain} = 2.5$gcm$^{-3}$, as described by Weingartner \& Draine (2001), was used to model the dust composition. This composition has been employed succesfully in the modeling of the Butterfly Star (Wolf et al. 2003) and CB26 (Sauter, Wolf et al. 2009). We extrapolated the complex refraction indices between 2mm${<}\lambda{<}4$mm to cover all available observational data. Because crystalline graphite is higly anisotropic, it is necessary to treat the two possible alignments independently using the so-called ''$\frac{1}{3}-\frac{2}{3}$'' approximation

\begin{equation}
Q = \frac{1}{3} Q(\epsilon_\parallel) + \frac{2}{3}Q(\epsilon_\perp).
\end{equation}
This approach explains the dependency of an arbitrary optical property $Q$ on the components of the dielectric tensor $\epsilon_\parallel$ and $\epsilon_\perp$ parallel and perpendicular to the symmetry axis and reproduces the observed interstellar extinction curve (Draine \& Malhotra 1993). We follow Wolf \& Voshchinnikov (2004) by calculating the optical properties of spherical particles with the subprogram \texttt{miex} of the \texttt{MC3D} package using Mie's solution and empirically measured indices of refraction.

\section{Parameter study}

\begin{table}
\caption{Canonical models of previous modeling work for HH 30 except W04. All parameters marked with the subscript $f$ were fixed or taken from previous publications in the corresponding study. Numbers marked with $\gtrsim$ are denoted as lower limits by the authors. W02 split luminosity into two contributions from the star and viscous dissipation. The dust masses assume a mass ratio of 1:100 between dust and gas.}
\centering
\begin{tabular}{lllll}
 & B96 & C01 & W02 & P06 \\
\hline
\hline

$\alpha$       & 2.2 & $2.367_f$ & $2.25_f$ & $2.2_f$ \\
$\beta$        & $1.45_f$ & 1.289 & 1.25 & $1.25_f$ \\
$R_{\rm{in}}$ [AU] & $0.5_f$ & 0.07 & 0.03 &  \\
$R_{\rm{out}}$ [AU] & ${\gtrsim}250_f$  & $200_f$ & $200_f$ & $145\pm20 $ \\
$h_{\rm{100}}$ [AU] & 15.5 & $15_f$ & 17 & $15.5_f$ \\
$m_{\rm{dust}} [10^{-5}M_\odot]$ & $6.0_f$ & $\gtrsim 0.67$ & $\gtrsim 1.5$ & $(2.7\pm0.4)$ \\
$L_* [L_\odot]$ & 1.0 & 0.2\ldots 0.9& 0.2 (+0.04) & $0.2_f$ \\
$T_{\rm{eff}}$ [K] & & 3000 & $3500_f$ & $3700_f$ (M1)\\
$i [^\circ]$ & 82.5 & $84_f$ & $84_f$ & $81\pm 3$ \\
\hline
\end{tabular}
\label{tab:previous}
\end{table}

The basis of understanding a phenomenon is the ability to predict the outcome of observations by a model. A parameter study consists of three subsequent steps that may be iterated until sufficient agreement between predictions and observation is reached. As a first step a particular model and a corresponding parameter space must be specified. In the following step a parameter set $\mathbf{a}_*$ is identified, which reproduces the observational data best with the assumed model. Finally an error bound $\Delta\mathbf{a}$ for this best-fit parameter set is characterized to evaluate the quality of the model. 

Finding the most likely parameter set is equivalent to locating the minimum of the chi-square distribution

\begin{equation}
\chi^2(\mathbf{a}) = \sum_{i=1}^N \frac{(\mu_i(\mathbf{a}) - y_i)^2}{\sigma_i^2}
\label{chi2}
\end{equation}
derived from $N$ measured quantities $(x_i{\pm}\Delta x_i, y_i{\pm}\Delta y_i)$ and their corresponding model predictions $\mu_i(\mathbf{a})$, e.g. the SED as pairs of wavelength and flux $(\lambda_i{\pm}\Delta\lambda_i,F_i{\pm}\Delta F_i)$. The error is given by

\begin{equation}
\sigma_i^2 = \Delta y_i^2 + \left[\frac{\partial\mu_i}{\partial x_i}(\mathbf{a})\cdot\Delta x_i\right]^2,
\end{equation}
but the second term can be neglected if $|\partial\mu_i/\partial x_i\cdot\Delta x_i|\ll|\Delta y_i|$, as is true most of the time. A straightforward method to identify the global minimum is to calculate all predictions on a fine grid and compare them with the observational data. This exhaustive search is not feasible in high-dimensional parameter spaces because it is not trivial to specify a sensible grid resolution \emph{a priori} and only a small sample of predictions can be calculated on a fine grid in a practical timeframe. Therefore one can only reduce the likelihood to find a better parameter set in most high-dimensional optimizations. If discrepancies between predictions and observations persist, the given model should be scrutinized and refined. This model review often introduces additional degrees of freedom and this increase in dimensionality may lead to complications and degeneracies during the optimization process, see Sect. \ref{implementation} for details of the refinements we implemented for this parameter study. HH 30 has already been modeled with similar models and less ample datasets (i.e. B96; C01; W02;  W04; P06), see Table \ref{tab:previous} for canonical models of these studies.

\subsection{Simulated annealing}
\label{sim_annealing}

\begin{table}[b!]
\caption{Subspace for the parameter study}   
\centering                     
\begin{tabular}{l l l l}       
\hline\hline                   
parameter & min & max \\       
\hline                         
   $\eta$            & 0.0 & 1.0 \\
   $\alpha$          & 2.0 & 3.0 \\
   $\beta$           & 0.5 & 2.0 \\
   $R_{\rm{in}}$ [AU]     &  0.1 & 70 \\
   $R_{\rm{att}}$ [AU]    & 0.1 & 70 \\
   $R_{\rm{out}}$ [AU]    &  100 & 500 \\
   $h_{\rm{100}}$ [AU]    &  5 & 20 \\
   $m_{\rm{dust}}$ [$M_\odot$] &  $1\cdot 10^{-6}$ & $2\cdot 10^{-2}$ \\
   $T_{\rm{eff}}$ [K]     &  3000 & 4000\\
   $L_*$ [$L_\odot$] &  0.1 & 1.0 \\
   $a_{\rm{min}}$ [$\mu$m] & 0.001  & 0.1 \\
   $a_{\rm{max}}$ [$\mu$m] & 0.125  & 65536 \\
   $d$               & 2.5 & 4.0 \\
   $ i [^\circ]$     & 79 & 86 \\
\hline                                   
\end{tabular}
\label{tab:para_subspace}
\end{table}

We used \emph{simulated annealing} (SA) to search for the best fit $\mathbf{a}_*$ (Kirkpatrick et al. 1983), a Markov Chain Monte Carlo method (MCMC) similar to the Metropolis-Hastings (MH) algorithm (Metropolis et al. 1953, Hastings 1970). This method has some specific advantages for high-dimensionality optimization because no gradients need to be calculated, local minima can be overcome intrinsically and points near optima will be evaluated with a higher probability. However, no upper bound for the step count to reach the global optimum can be given and a reasonable criterion for chain abortion must be defined.

We denote in the following the sequence of all propositions, i.e. rejected and accepted, by $(\mathbf{a}_n)$ and the canonical Markov chain of all accepted states by $(\overline{\mathbf{a}}_n)$. The random walk through the parameter space $P=I_1\times\ldots\times I_D\subset\mathbb{R}^D$ is generated by starting at an arbitrary point $\mathbf{a}_0\in P$ and subsequently sampling steps $\mathbf{\Delta a}$ from a \emph{transition probability} T. We chose a Gaussian distribution

\begin{equation}
T(\mathbf{\Delta a}) \sim \prod_{j=1}^D \exp\left[-\frac{\Delta a_j^2}{2\beta_j^2} \right]
\label{proposition}
\end{equation}
with predefined step widths $\beta_j$ for every dimension.  The exact form is not important as long as the parameter space can be exhausted in principle and $\overline{\mathbf{a}}_n+\Delta\mathbf{a}\in P$. Optimal step sizes cannot be determined a priori because no information is available on the distribution under investigation, and they have to be adapted along the way. After calculating all predictions $\mu_i(\mathbf{a}_{n+1})$ at the proposed next location in parameter space, the \emph{acceptance probability}
\begin{equation}
A = \min\left\{ 1, \exp\left[-\frac{\Delta\,\chi^2}{\tau}\right]\right\}
\label{MH_decision}
\end{equation}
is evaluated by calculating the difference $\Delta\chi^2 = \chi^2_{n+1} - \overline{\chi}^2_{n}$ between proposed and actual position in parameter space with eq. (\ref{chi2}), $\tau$ is the effective temperature of the "annealing" process. A uniformly distributed random number $u\in]0,1[$ is generated and if $u<A$, the step is taken, otherwise the proposition is rejected and a new step $\Delta\mathbf{a}$ is generated. This decision is central to the MH algorithm and generates the Markov chain through phase space by taking every downhill step while still enabling the escape from a local valley depending on the actual temperature $\tau$. 

Most implementations of SA use a monotonic cooling schedule, but these schemes contain several parameters that must be adjusted empirically by running the optimization several times. Because the calculation of predictions $\mu_i(\mathbf{a})$ in our study is very time consuming, a self-regulating algorithm is preferable. We therefore used a non-monotonic cooling scheme by setting the temperature for the $n$-th step to 

\begin{equation}
\tau_n = \theta\left[\overline{\chi}^2_n - \chi^2_*\right]
\end{equation}
with a fixed parameter $\theta$, the $\overline{\chi}^2_n$ of the previously accepted step, and the $\chi^2_*$ of the global optimum (Bohachevsky et al. 1986; Locatelli 2000). The latter value is typically unknown and one has to resort to a lower bound or sensible approximation for $\chi^2_*$ like the $\chi^2$ of the current best-fit of an individual run. The parameter $\theta\lesssim 1$ has to be fixed empirically to maintain the temperature in the order of typical fluctuations $\Delta\chi^2$ encountered during optimization. 

We defined the local acceptance ratio of the chain using the sequence $(\kappa_n)\in\left\{0,1\right\}$ of accepted and rejected steps by
\begin{equation}
\xi_n = \frac{1}{l}\sum_{m=n-l+1}^n \kappa_m
\end{equation}
for some constant lookback length $l<n$. To remain in the vicinity of the optimal acceptance ratio $\hat{\xi}{\sim}0.234$  (Roberts et al. 1997), we implemented a control loop that adjusts the step widths $\beta_j$ of the transition density (\ref{proposition}) on the fly.  The algorithm analyzes the last $l$ positions of the chain and calculates the moduli of relative changes between the proposed parameter vector $\mathbf{a}_n$ and the previously accepted parameter set $\overline{\mathbf{a}}_{n-1}$:

\begin{equation}
c_{n,j} = \left|\frac{a_{n,j} - \overline{a}_{n-1,j}}{\overline{a}_{n-1,j}}\right|.
\end{equation}
The index $j\in\left\{1,..,D\right\}$ enumerates here the parameter dimensions. By separately summing up the relative change vectors $\mathbf{c}_n\in\mathbb{R}^D$ in accepted and rejected proposals two vectors $\mathbf{g}_n$ and $\mathbf{b}_n$ can be derived, which describes the impact of good and bad decisions:

\begin{equation}
\mathbf{g}_n = \sum_{m=n-l+1}^n \kappa_m \mathbf{c}_m,\quad  \mathbf{b}_n = \sum_{m=n-l+1}^n (1-\kappa_m) \mathbf{c}_m.
\end{equation}
If $\xi_n>\hat{\xi}$ the parameter with the smallest component in $\mathbf{g}_n$ is modified to encourage riskier proposals. In the opposite case the parameter with the largest component in $\mathbf{b}_n$ is reduced to induce a more conservative approach. In either case the corresponding stepping size $\beta_r$ is adjusted proportionally by multiplying with $(1 - \hat{\xi}) + \xi_n$. To avoid \emph{freezing} of individual parameters during the run, it is recommended to define a reasonable lower bound for every $\beta_j$, i.e. enforce $\beta_j\geq\delta|I_j|$ with $\delta\ll 1$.

\subsection{Calculation and combination of $\chi^2$}
\label{chi2_calc}
There are three fundamentally different datasets available for our study: the SED, resolved images, and interferometric visibilities. The methods to calculate the corresponding\, $\chi^2_k$ differ and we will outline the involved procedures. 

The computation of $\chi^2_{\rm{SED}}$ is straightforward because the independent measurements can just be entered into eq. (\ref{chi2}). 

To compare observed and simulated brightness distributions the synthetic images must be scaled to the pixel scale of the original image, convolved with the PSF of the corresponding instrument setup, and then registered with the observed image. There are plenty of image registration algorithms, in our case the images only needed translation, therefore it was sufficient to evaluate the correlation integral or match the flux maxima.

To calculate the deviation $\chi^2_{\rm{visib}}$ for the observed visibilities $V_i$, one has first to generate visibilities $W_i$ from a simulated image by Fourier transformation. To achieve sufficient resolution in the uv plane, the image must be padded with zeroes prior to using the FFT as size and count of pixels of an $M\times N$ array are connected by
\begin{equation}
M\Delta x\Delta u = N\Delta y\Delta v = 1.
\end{equation}
The complex visibilities $W_i$ are then interpolated from the resulting array using the actual antenna tracks in the uv plane. Because the real and imaginary parts were measured independently with the PdBI, we used the expression

\begin{equation}
\chi^2_{\rm{visib}} = \sum_i \frac{|\Re(V_i) - \Re(W_i)|^2 + |\Im(V_i) - \Im(W_i)|^2}{\sigma_i^2}.
\label{chi2_visib}
\end{equation}
The center of the object does not lie exactly in the origin of the coordinate system during observation, equation (\ref{chi2_visib}) must hence be minimized by translating the visibilities in the uv plane using the expression 

\begin{equation}
W_i' = W_i\cdot \exp\left[-2\pi i(xu_i + yv_i)\right]
\end{equation}
to find the best translation vector $(x_0,y_0)$.

It is often necessary to combine $\chi^2_k$ derived from independent observables. One straightforward method is linear combination of the individual values

\begin{equation} 
\chi^2_{\rm{total}} = \sum_k g_k \,\chi^2_k,
\end{equation}
which introduces an arbitrary weighting through the coefficients $g_k$. Unfortunately, the signal-to-noise ratio (S/N) or even the noise itself can differ strongly between observables and it is not clear how to choose the coefficients $g_k$. 

Confronted with this problem we devised a \emph{staggered} Markov chain. Instead of performing all calculations at once to evaluate all $\chi^2_k$ and directly combine them to decide to accept a proposed configuration or reject it, we performed an MH decision after each individual calculation of $\chi^2_k$. We hence needed to keep track of individual temperatures $\tau_k$ for every dataset that is updated as described in the previous paragraph. By only accepting new configurations after all individual $\chi^2_k$ steps were completed succesfully, the algorithm effectively optimizes all observational data sets simultaneously \emph{without} weighting them. An added advantage of this procedure is that there is less computing time wasted on uninteresting parameter sets, especially if less time-consuming calculations are performed first. In any case, after performing the MCMC optimization it is necessary to choose the weights $g_k$ to fix the metric of $\chi^2_{\rm{total}}$ by analyzing the calculated samples.

\subsection{Confidence interval}
\label{confidence}

Localizing a best-fit parameter set $\mathbf{a}_*$ is the first step of the fitting process. The second step is estimating a confidence region surrounding this optimum, which can be established by sampling the $\chi^2$-distribution and locating the confidence boundary given by 
\begin{equation}
\chi^2_{\rm{conf}} = \chi^2_* + \Delta\chi^2_{\rm{conf}}. 
\end{equation}

Sampling of an unknown distribution was the prime motivation behind the MH-algorithm, therefore no major code change is necessary to estimate error bounds. 
To identify an appropriate $\Delta\chi^2_{\rm{conf}}$ the \emph{reduced} merit function
\begin{equation}
\chi^2_{\rm{red}} = \frac{\chi^2}{K}
\end{equation}
is often used, where $K$ is the number of degrees of freedom of the model. Unfortunately, $\chi^2_{\rm{red}}$ cannot be employed for non-linear models because $K$ is unknown and can change depending on the position in parameter space (Andrae et al. 2010). Additionally, the S/N can differ significantly between observables which leaves the choice of $\Delta\chi^2_{\rm{conf}}$ to the practicioner's discretion. After choosing a sensible $\Delta\chi^2_{\rm{conf}}$, the vicinity can be sampled by several Markov chains starting from $\mathbf{a}_*$ with a constant temperature $\tau$ to probe the Boltzmann-distribution $\exp\left[-\chi^2/\tau\right]$. The constant temperature has to be adjusted to prevent the chains from leaving the vicinity of $\mathbf{a}_*$ too fast, e.g. $\tau\lesssim\Delta\chi^2_{\rm{conf}}$. To bootstrap this procedure one can inspect the findings of previous Markov chains used for SA. Otherwise, the region has to be explored first by using an educated guess for the temperature. After sufficient sampling, the error margin $\Delta\mathbf{a}_*$ can be deduced by extracting the minimal and maximal components of all sampled parameter vectors $\mathbf{a}_i$ with $\chi^2_i<\chi^2_{\rm{conf}}$. Because the confidence region is normally distorted by degeneracies and non-linearity of the underlying model, the margin $\Delta\mathbf{a}_*$ is composed of non-centered intervals containing the components of $\mathbf{a}_*$, see Table \ref{tab:best_fit}. The reader is kindly reminded that we cannot exclude with certainty that there might be a similar or better solution beyond these error intervals. This would require a global search of the 14-dimensional parameter space on a sufficiently refined grid, a task not feasible with currently available computing resources.

\subsection{Implementation}
\label{implementation}

We started our model fitting with the high-resolution interferometric continuum images obtained with the PdBI described in Sect. \ref{pdbi_imaging}. The dip between the two maxima in Fig.~\ref{hh30_sigma} is consistent with an inner hole in the optically thin case (G08). Molecular line observations made with the PdBI show a single gaseous disk in Keplerian rotation (P06), excluding the possibility that each peak of the continuum image is actually an individual disk
revolving around one of the components of a wide binary. In the optically thin case, the total dust mass $m_{\rm{dust}}$ and flux $F_\lambda$ are connected by
\begin{equation}
m_{\rm{dust}} \kappa_\lambda = \frac{F_\lambda d^2}{B_\lambda(\left<T_{\rm{dust}}\right>)}
\label{dust_mass}
\end{equation}
with the distance $d$ to the object and the mean temperature $\left<T_{\rm{dust}} \right>$ of the dust. This equation can only be solved for $m_{\rm{dust}}$ if the dust opacity $\kappa_\lambda$ is known, hence normally only the left-hand side of equation (\ref{dust_mass}) can be evaluated (e.g. Reipurth et al. 1993).  By choosing a particular dust model, the total dust mass becomes fixed, thereby eliminating one modeling parameter in the optically thin case. The lack of detectable emission in the interferometric image beyond $R{\gtrsim}130$AU is inconsistent with the extension deduced from optical and near-infrared images, as can be seen by superposing observations made with the HST (G08). It indicates a significant change in dust emissivity or drop below the detection limit at this distance.

With this information in mind, we implemented a simple flared disk model, using the parameters $\alpha$, $\beta$, $R_{\rm{in}}$, $R_{\rm{out}}$, $h_{\rm{100}}$, $m_{\rm{dust}}$, $T_{\rm{eff}}$, $L_*$ and $i$ with the standard MRN dust composition because there was no need for more free parameters at this initial stage of the investigation. By allowing for a varying dust mass the possibility of optical depth effects was acknowledged. We conducted a grid search on the corresponding subset of Table \ref{tab:para_subspace} by dividing the nine intervals of the implemented parameters into $3\ldots 5$ parts. For every point on the grid the temperature distribution and a raytraced image at $\lambda =1.3$mm was generated, then convolved using a Gaussian PSF with FWHM extension $0.59^{\prime\prime}{\times}0.32^{\prime\prime}$ at $PA=22^\circ$ as described in G08 and scaled to fit the angular resolution of the deconvolved mm map. The shapes of measured and synthesized brightness distributions were compared by normalizing to a maximum flux of 1, registering by correlation, and calculating $\chi^2_{\rm{shape}}$ by subtracting and squaring. This approach led us to the following intermediate results:

\begin{enumerate}
\item The interferometric image can be explained by a flared disk with an inner hole of $R_{\rm{in}}=(55{\pm}15)$AU and an outer radius of $R_{\rm{out}}= (130{\pm}30)$AU.
\item The vertical geometry of the system can be best reproduced by a thin, ring-like disk with a flat (i.e. constant) surface density distribution.
\end{enumerate}
As a next step, we investigated the SED of best-fit models found by the grid search. To exclude variability of the star and unknown foreground extinction from the modeling process, only data points with $\lambda>5\mu$m were used to calculate $\chi^2_{\rm{SED}}$. Soon it became clear that a flared disk model with a large hole and simple MRN dust cannot reproduce the observed fluxes and spectral indices, either in the MIR or the mm domain. To counter the missing flux in the MIR, we introduced a depletion zone parametrized by an attenuation factor and radius as described in Sect. \ref{disk_model}, effectively adding warm dust to the inner part of the disk. Additionally, we allowed for a variable grain size distribution as described in Sect. \ref{size_distr} to include larger dust grains in the mix to improve the spectral index and total flux, especially at longer wavelengths. Expansion of the parameter space by the new parameters $\eta$, $R_{\rm{att}}$, $a_{\rm{min}}, a_{\rm{max}}$, and $d$ rendered a grid search ineffective and led to implementation of SA as described in chapter \ref{sim_annealing} to search for the global optimum in the parameter subspace defined by Table \ref{tab:para_subspace}.

Four optimization runs were performed with this technique using different subsets of the observations described in Sect. \ref{observations}. They resulted in four best-fit parameter sets that we will refer to as model A, B, C, and D in the following. Only the NIR images taken with NICMOS/HST where used in the optimization run for model A, see Fig.~\ref{fig:nic2_images}. Model B is based on all available SED observations with $\lambda{\gtrsim}5\mu$m, see Fig.~\ref{hh30_sed}. The same data set was employed to find Model C, but with the restriction $\eta{=}1$ to implement the limiting case of a simple flared disk. Model D is the best fit for simultaneous optimization of SED data and mm image from G08, see Fig.~\ref{hh30_sigma}. For model A, we started eight Markov chains to fit the image observed with NICMOS using the F204M filter and applying in principle the same procedure implemented for the interferometric image. We rotated the original image by $-32^\circ$, cut out a region of $4"\times4"$ centered on our object, re-sized to $400\times400$ pixels and smoothed with bilinear interpolation to reduce artifacts from the rotation, see Fig.~\ref{fig:nic2_images}. Synthetic scattered light images were generated, then convolved with the corresponding \texttt{Tiny Tim} PSF, and finally registered and compared to the observed image. To eliminate the unknown foreground extinction and variation in luminosity, we set the error and peak of simulated and reference image to unity for calculating $\chi^2_{\rm{F204M}}$, hence optimizing only the shape. Some fair matches for the F204M filter observation were found after a few hundred steps. We then calculated the corresponding images at the F110W and F160W filter wavelengths for the best 50 images found and searched for quantitative matches in all three filters by dereddening the simulated fluxes to match the observation in addition to the already fitted shape. The parameter set and $\chi^2_{\rm{F204M}}$ of the resulting model A are listed in Table \ref{tab:best_fit}, its optical properties can be found in Table \ref{tab:modelA_prop}. This parameter set should not be understood as an overall best-fit, but just as the best model found by our chains. As W04 discussed extensively, the model is too degenerate to pinpoint a location in parameter space given scattered light images alone. Especially parameters describing the inner disk, i.e. $R_{\rm{in}}$, $R_{\rm{att}}$, and $\eta$, may not be constrained at all by the scattered light images.

The results of the NIR data optimization suggested that

\begin{figure}
  \centering
  \includegraphics[width=0.5\textwidth]{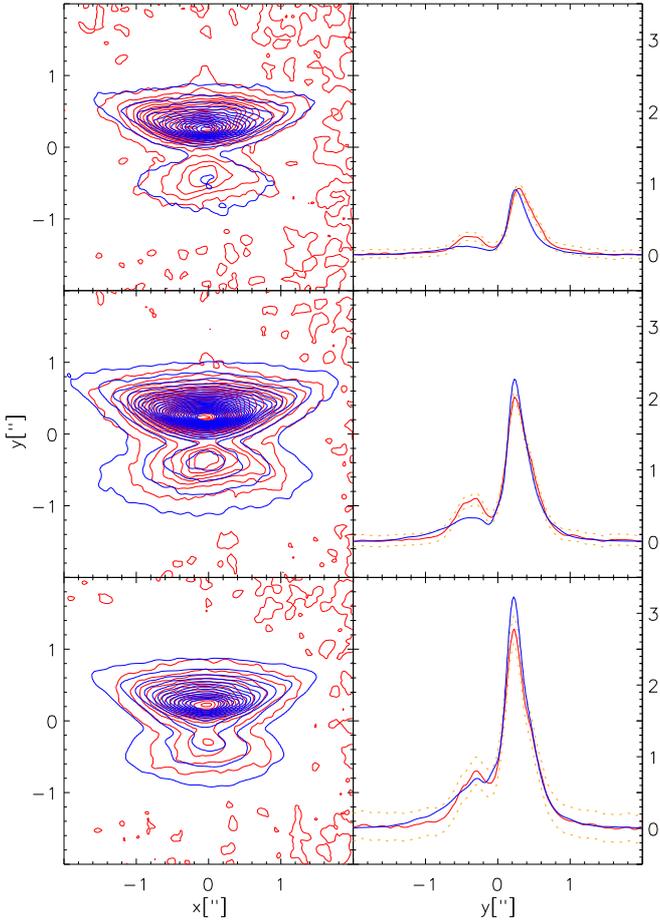}
  \caption{Contour plots and vertical cuts along the symmetry axis of the HST/NICMOS observations (red) and the best-fit model A (blue). The contours in the left column and the dotted intervals in the right column correspond to 1$\sigma$.}
  \label{fig:scatter_contour}
\end{figure}

\begin{enumerate}
\item Small dust grains with $a_{\rm{max}}\lesssim 2\mu$m are sufficient to reproduce the observed flux, optical thickness and chromaticity.
\item Observations in scattered light are consistent with the postulated depletion zone. The specifics of the inner region do not change the overall
appearance, as proposed by W04.
\item Parameters influencing the vertical geometry ($h_{100}, \beta$) are better constrained than radial parameters ($R_{\rm{in}}$, $R_{\rm{att}}$, and $R_{\rm{out}}$).
\item Contrasts in vertical cuts between the secondary maximum and enclosed minimum in the synthetic scatter images are systematically too low by a factor of ${\sim}2$, see Fig.~\ref{fig:scatter_contour}.
\end{enumerate}

For model B, we fitted the SED using 32 individual MCMC runs started from randomly selected points and propagated freely for ${\sim}1000$ steps through the parameter subspace. Several Markov chains found parameter sets with low $\chi^2_{\rm{SED}}$, the best parameter set of this run is shown in Table \ref{tab:best_fit} as model B. The low spectral index $\beta_{\rm{mm}}{\sim}0.4$ indicates the presence of large dust grains, i.e. $a_{\rm{max}}\gtrsim 1$mm, and a grain size exponent $d<3.5$ (Draine 2006). However, the models investigated in this parameter range often lack flux and exhibit a very different spectral index in the MIR than observed. Nevertheless, these MCMC runs gave us more hints:

\begin{enumerate}
\item Large dust grain sizes $a_{\rm{max}}\gtrsim1$mm are necessary to explain the observed flux, especially at longer wavelengths.
\item The flux and spectral index at mm-wavelengths can be reproduced with a dust exponent $d{\sim}3$ and $a_{\rm{max}}\gtrsim 8$mm.
\item No synthetic SED fits the observations at short and long wavelengths simultaneously.
\item The geometry of the object is only weakly constrained by the SED, i.e. the model is degenerate.
\end{enumerate}

It is noteworthy that the best solution and its local confidence interval suggest a change in dust density by a factor of ${\gtrsim}10$ at a distance of ${\sim}45$AU, thereby dividing the system into an inner and outer disk, see Fig.~\ref{density_diagram}.

To ensure that this finding was not biased by introducing the depleted zone, we also conducted an optimization without this feature by fixing $\eta{=}1$, which resulted in model C. We set up 24 Markov chains to search for solutions with this simple flared disk model and checked their appearance at 1.3mm.  After ${\sim}1000$ steps, several Markov chains found parameter sets with $\chi^2_{\rm{SED}}$ comparable to, but not better than, model B. 
The best parameter set of this run is given in Table 5. The 1.3mm map generated using these parameters is representative for the parameter sets for this run and is centrally peaked, showing poor agreement with the observed brightness profile, see Fig.~\ref{hh30_sigma}.

The most apparent discrepancy between models A and B is the different grain size distribution and thus the total dust mass. In an attempt to reconcile all approaches, we implemented the simultaneous optimization of SED and brightness map at $\lambda=1.3$mm with a staggered Markov chain as described in Sect. \ref{chi2_calc}. We started 24 Markov chains and let them run for ${\sim}1000$ steps to search for a unified model. The parameter set listed in Table \ref{tab:best_fit} as model D is the best compromise. The analysis of this run gave us additional insights:

\begin{enumerate}
\item Combination of independent datasets stall optimization, i.e. the Markov chains need longer to find a compromise.
\item Staggered optimization reduces model degeneracies, e.g. the outer radius $R_{\rm{out}}$ moved to values $\lesssim 200$AU by including the mm image in the optimization.
\end{enumerate}

The latter effect was expected from the results obtained by our initial grid search, but solutions with $R_{\rm{out}}\gtrsim 200$AU are still difficult to exclude with certainty using only direct comparison of brightness maps at mm-wavelengths. To circumvent possible artifacts introduced by the CLEAN procedure we implemented direct fitting of the visibilities in the uv plane. Analysis of all ${\sim}25000$ synthesized mm images showed that the measured visibilities can be best explained with $R_{\rm{out}}=(135{\pm}15)$AU. Unfortunately, these models have difficulties reproducing the flux in the MIR and we concluded that further optimization should be continued in a follow-up study when additional high-resolution observations in the (sub)mm domain are available that reveal the spatial dependence of the spectral index $\beta_{\rm{mm}}$.

\begin{figure}
  \centering
  \includegraphics[width=0.5\textwidth]{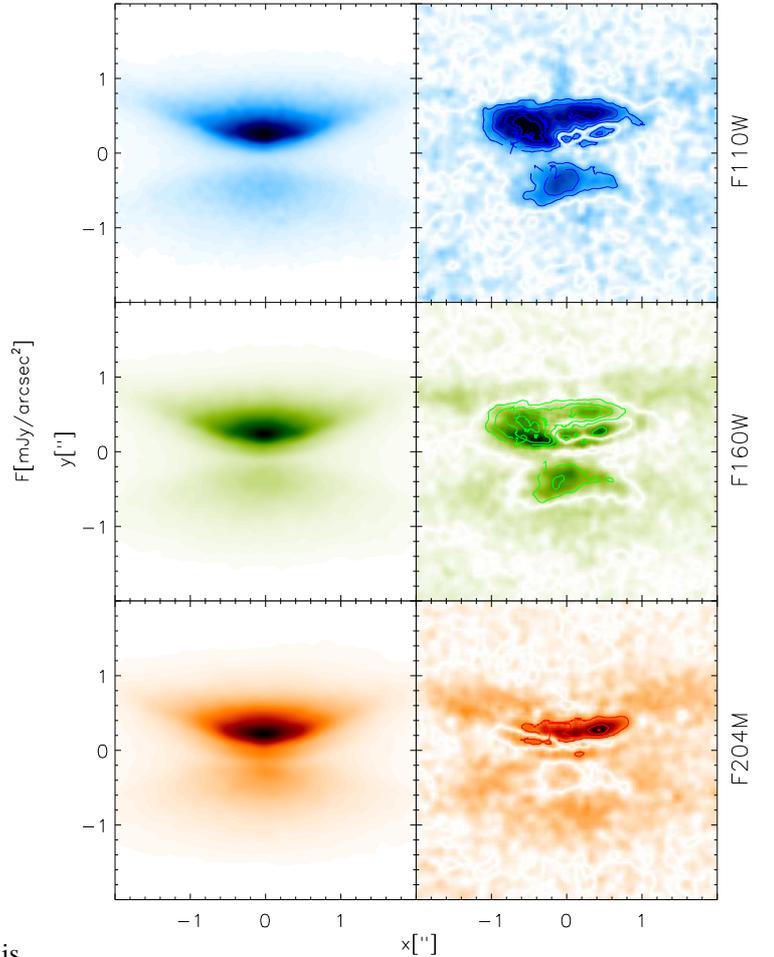}
  \caption{Simulated images of model A in the NIR and the deviations from observations with NICMOS with overlayed contour lines spaced at 1$\sigma$. As expected, the differences are lowest in the optimized F204M image and increase toward shorter wavelengths.}
  \label{fig:scatter_sub}
\end{figure}

\begin{table}
\caption{Dust opacity $\kappa$, albedo $a$ and asymmetry factor $g=\langle\cos\theta\rangle$ for model A. Values in parentheses are from the best-fit model of C01.}
\centering
\begin{tabular}{l l l l l}
\hline
\hline
Filter & $\lambda_c$ [$\mu$m] & $\kappa$ [cm$^2$ g$^{-1}$] & a & g \\
\hline
F110W & 1.14 & 12333 (12100) & 0.68 (0.48) & 0.55 (0.52) \\
F160W & 1.61 & 9731   (8790) & 0.70 (0.48) & 0.52 (0.49) \\
F204M & 2.08 & 8234   (6800) & 0.71 (0.46) & 0.50 (0.46) \\
\hline
\end{tabular}
\label{tab:modelA_prop}
\end{table}

\begin{table}
\caption{Best-fit and deduced parameters found by optimization runs for the NIR images (A), SED data only (B), SED data only for the simple flared disk model with $\eta{=}1$ (C), and mm-map/SED combined (D). The confidence intervals of model B were deduced by the procedure described in Sect. 4.3 and are discussed in Sect. 5.2.}    
\centering                     
\begin{tabular}{l l l l l} 
\hline\hline                   
model & A  & B & C & D \\      
\hline                         
 $\eta$                             &    0.019        & $0.029^{+0.026}_{-0.002}$  & 1      & 0.03     \parbox[0pt][1.5em][c]{0cm}{} \\
 $\alpha$                           &    2.45         & $2.27^{+0.24}_{-0.10}$     & 2.39   & 2.3    \parbox[0pt][1.5em][c]{0cm}{} \\
 $\beta$                            &    1.09         & $1.18^{+0.03}_{-0.07}$     & 1.21   & 1.16   \parbox[0pt][1.5em][c]{0cm}{} \\
 $p$                                &    1.36         & 1.09                       & 1.17   & 1.14   \parbox[0pt][1.5em][c]{0cm}{} \\
 $R_{\rm{in}}$ [AU]                 &    2.1          & $0.13^{+0.83}_{-0.03}$     & 0.05   & 0.56   \parbox[0pt][1.5em][c]{0cm}{} \\
 $R_{\rm{att}}$ [AU]                &   59.6          & $47.0^{+1.4}_{-5.2}$       & na     & 44.4   \parbox[0pt][1.5em][c]{0cm}{} \\
 $R_{\rm{out}}$ [AU]                &  455            & $471^{+24}_{-346}$         & 212    & 175    \parbox[0pt][1.5em][c]{0cm}{} \\
 $h_{\rm{100}}$ [AU]                &   14.8          & $14.9^{+0.3}_{-1.4}$       & 14.8   & 14.7   \parbox[0pt][1.5em][c]{0cm}{} \\
 $m_{\rm{dust}}$ [$10^{-5}M_\odot$] &    0.734        & $7.3^{+1.6}_{-3.1}$        & 9.47   & $5.04$ \parbox[0pt][1.5em][c]{0cm}{} \\
 $m_{\rm{in}}/m_{\rm{out}}$         &    0.09         &  0.11                      & na     & 0.15   \parbox[0pt][1.5em][c]{0cm}{} \\
 $T_{\rm{eff}}$ [K]                 &    3200         &  $3200^{+400}_{-200}$      & 4000   & 3300   \parbox[0pt][1.5em][c]{0cm}{} \\
 $L_*$ [$L_\odot$]                  &    0.6          &  $0.4^{+0.11}_{-0.06}$     & 0.38   & 0.48   \parbox[0pt][1.5em][c]{0cm}{} \\
 $a_{\rm{min}}$ [$\mu$m]            &    0.006        &  $0.009^{+0.001}_{-0.008}$ & 0.005  & 0.003  \parbox[0pt][1.5em][c]{0cm}{} \\
 $a_{\rm{max}}$ [$\mu$m]            &    1.87         &  $18800^{+7800}_{-7200}$   & 7130   & 20200  \parbox[0pt][1.5em][c]{0cm}{} \\
 $d$                                &    3.4          &  $3.7^{+0.06}_{-0.18}$     & 3.35   & 3.58   \parbox[0pt][1.5em][c]{0cm}{} \\
 $i [^\circ]$                       &   83.6          &  $83.2^{+2.1}_{-1.5}$      & 80.1   & 85.1   \parbox[0pt][1.5em][c]{0cm}{} \\
\hline                                   
$\chi^2_{\rm{F204M}}$               &  49             &  586                  & 1079   & 781    \parbox[0pt][1.5em][c]{0cm}{} \\
$\chi^2_{\rm{SED}}$                 &  $8.2\cdot10^7$ &  12884                & 59130  & 41471  \parbox[0pt][1.5em][c]{0cm}{} \\
$\chi^2_{\rm{1.3mm}}$               &  177287         &  74416                & 77438  & 55239  \parbox[0pt][1.5em][c]{0cm}{} \\
\hline
\end{tabular}
\label{tab:best_fit}
\end{table}

\section{Results}

\begin{figure}
  \centering
  \includegraphics[width=0.5\textwidth]{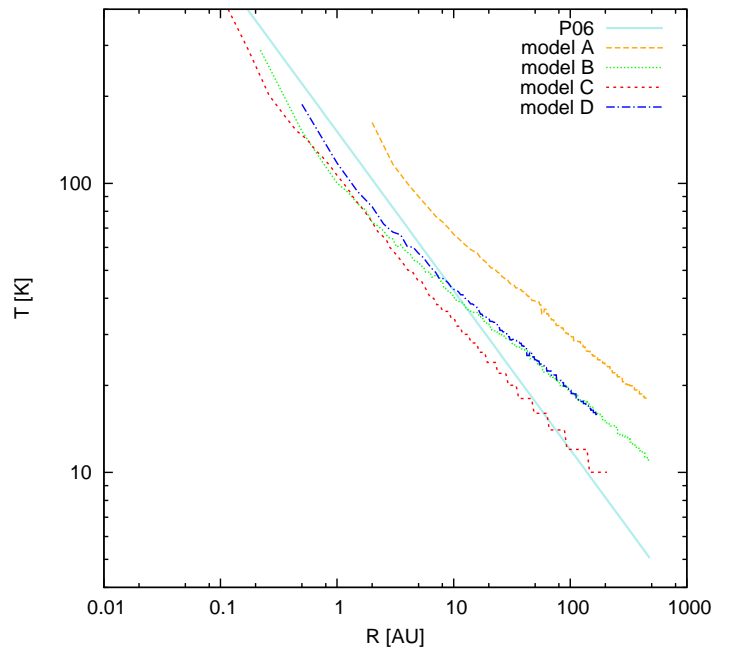}
  \caption{Temperature in the mid-plane as predicted by our best-fit models and the best-fit power law of P06 deduced from $^{13}$CO observations.}
  \label{fig:best_temp}
\end{figure}

The models mentioned in the preceding section are presented in Table \ref{tab:best_fit} for direct comparison. The parameters of previous studies are compiled in Table \ref{tab:previous} and agree well with our results. This is not self-evident because our parameter study was unbiased and employed an extended disk model with 14 parameters prone to degeneracies. 

The radial temperature distributions in the mid-plane of our models are compared in Fig.~\ref{fig:best_temp} with the power law deduced from $^{13}$CO observations by P06

\begin{equation}
T(R) = T_{100} \left(\frac{R}{100}\right)^{-q}
\label{power_law}
\end{equation}
with  $T_{\rm{100}} = (12{\pm}1)$K and $q=(0.55{\pm}0.07)$. Vertical cuts along the symmetry axis of the brightness maps at $\lambda=1.3$mm are depicted in Fig.~\ref{fig:mm_slices} in direct comparison with the observation by G08. In Fig.~\ref{fig:best_SED} synthetic SEDs for $5\mu$m$<\lambda<4$mm are overlayed on the observed fluxes. We discuss the individual properties of our best-fit models in the following sections. 

\subsection{Fitting of NIR observations}
Apart from small asymmetries due to uneven illumination and influence of the jet, model A appears to be a fair match to the NICMOS/HST observations, see Fig.~\ref{fig:scatter_contour} and \ref{fig:scatter_sub} for a direct comparison of simulated and measured scattered brightness distributions. However, it must be noted that we were not able to reproduce the contrast between the secondary maximum and the central minimum in the vertical cuts, as can be seen in the right column of Fig.~\ref{fig:scatter_contour}. This phenomenon has also been reported in previous studies using scattered NIR images (C01 and W04), indicating a systematic deviation of our model from reality. One possible explanation for this discrepancy might be the cone of low-velocity gas driven by the central jet that has been detected in molecular lines (P06). This gas stream can raise dust (Tambovtseva et al. 2008) that would substantially enhance scattering of photons from the central source in this region and hence contrast. Again, we decided not to increase the complexity of the model without additional observational data supporting this claim because other scenarios such as disk warps, shadowing, and asymmetric illumination by a binary (e.g. Tambovtseva et al. 2006) may cause the same effect.

The models of B96 and C01 are the counterparts of model A because these studies used optical and NIR observations as the basis for modeling. The most prominent difference is the outer radius $R_{\rm{out}}=455$AU, which appears to be much larger than corresponding values found by our predecessors. This discrepancy is not as large as it may appear at first glance because the best-fit models of previous studies that treated $R_{\rm{out}}$ as a free parameter, i.e. B96 and P06, yielded $R_{\rm{out}}{\sim}425$AU and ($420{\pm}20$)AU, respectively. The flux scattered by outer parts of our models with large radii is much lower than 1$\sigma$ and is not significant compared with observations, especially with the relatively high surface density exponent $p{\sim}1.4$ of model A. Therefore, we show in Figs.~\ref{fig:scatter_contour} and \ref{fig:scatter_sub} only the region with flux high enough to be  meaningfully compared with observations. Our modeling results suggest that NIR observations alone cannot yield an upper limit for $R_{\rm{out}}$, therefore the lower values found by previous studies are lower bounds and our proposed value should be treated with caution. 

The ratio of dust opacities for the used filter wavelengths of $\kappa_{\rm{F110W}}$:$\kappa_{\rm{F160W}}$:$\kappa_{\rm{F204M}}$=1.50:1.18:1.00 is comparable to the best-fit of C01 (1.37:1.29:1.00), as are most optical dust properties, see Table \ref{tab:modelA_prop}.

Despite the promising appearance of model A in the NIR, its SED and mm map are disappointing, see Figs.~\ref{fig:mm_slices} and \ref{fig:best_SED}. This model is missing some essential ingredient and its deviation from the observed temperature in the mid-plane as depicted in Fig.~\ref{fig:best_temp} underlines this fact.

\subsection{Fitting of the SED}
\begin{figure}
  \centering
  \includegraphics[width=0.5\textwidth]{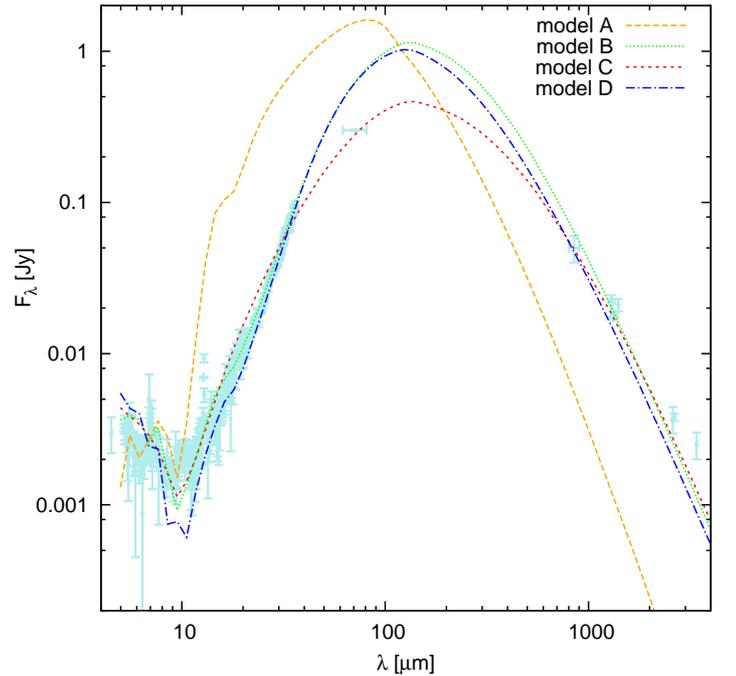}
  \caption{Direct comparison of the synthesized SEDs of our best-fit models with observations in the investigated wavelength domain from $5\mu$m to 3mm.}
\label{fig:best_SED}
\end{figure}

Model B predicts the SED quite well, see Fig.~\ref{fig:best_SED}, and comparison with model A uncovers the necessary ingredient: larger grains with radii $a\gtrsim1$mm. This result is congruent with the findings by W02, although these authors used an exponential cutoff for their grain size distribution. Unfortunately, model B cannot reproduce the NIR observations although the overall geometry of models A and B are very similar. The presence of larger grains in model B changes the optical properties of the dust to the point that this configuration is no longer a good match for the NIR observations. Nonetheless, these large grains are necessary to account for the observed fluxes and spectral indices at mm-wavelengths and model B is in fact a tradeoff because it is not able to reproduce the low spectral index $\beta_{\rm{mm}}{\sim}0.4$ and the observed fluxes in the mm domain with the same confidence as in the MIR. Contrary to intuition, no silicate features are visible in the model SED although the grain ensemble contains sufficiently small grains. This is a geometric effect caused by shadowing of the inner, hot disk because tilting by a few degrees unmasks the hidden 10$\mu$m bright region.

The slice through the mm map shows that the emission of model B at $\lambda=1.3$mm is too extended and has insufficient peak brightness and contrast. Models with lower dust exponents that are able to match the (sub)mm flux exist, but they cannot reproduce the well-sampled SED in the MIR domain. 

The radial temperature distribution in the mid-plane of model B follows for $R{\gtrsim}5$AU the power law (\ref{power_law}) with $T_{\rm{100}} = 19$K and $q=0.33$. The temperature varies from $T{\sim}290$K at the inner edge to $T{\sim}10$K in the outer regions with a mass average temperature of $\left<T_{\rm{dust}}\right>{\sim}18K$. Our model predicts slightly higher temperatures in the midplane than deduced from observations by P06, but is still a fair match (see Fig.~\ref{fig:best_temp}).

Effective temperature and luminosity are consistent with a contracting protostar of spectral class M3${\pm}2$ with $R_*{\sim}2R_\odot$ (Simon et al. 2000), but it is very difficult to constrain the properties of the central source and a binary, as proposed by Anglada et al. (2007) and G08, seems a viable option.

The surface density exponent $p{\sim}1.2$ is larger than expected for a thin accretion disk with $\alpha$-viscosity, i.e. eq. (\ref{eq:alpha_beta}) does not hold. This is no surprise because the postulated inner depletion zone must be generated by some additional physical process.

The calculation of confidence intervals was performed for model B as described in Sect. \ref{confidence} by sampling the $\chi^2$-distribution around $\mathbf{a}_B$ and is based on ${\sim}1000$ simulated SEDs in the vicinity of this parameter set. We chose $\Delta\chi^2_{\rm{conf}}{\sim}0.5\cdot\chi^2_B$ and deduced the asymmetric intervals given in Table \ref{tab:best_fit}. The reader is cautioned to not confound these confidence intervals with a classical $1\sigma$ error interval because the model is highly degenerated and consequently we are not dealing with independent parameters. Additionally, the selection of $\Delta\chi^2_{\rm{conf}}$ is based on the subjective quality of the fit to the observational data as $\chi_{\rm{SED,red}}^2\gg1$. Comparison of the confidence intervals with the parameter subspace defined in Table \ref{tab:para_subspace} shows that the upper interval of $R_{\rm{out}}$ has been clipped. As discussed in Sect. 5.1, we cannot give an upper constraint for this parameter with available observations and accordingly did not explore the confidence interval beyond 500AU. The parameters $T_{\rm{eff}}$ and $L_*$ describing the source might also be influenced by clipping because their intervals almost cover the given subspace. The error bounds reflect the degeneracy in the vicinity of our best-fit model B and remind us of the difficulties of constraining a model in a high-dimensional parameter space using only SED data.

\subsection{Fitting of the SED with a simple flared disk}

\begin{figure}
  \centering
  \includegraphics[width=0.5\textwidth]{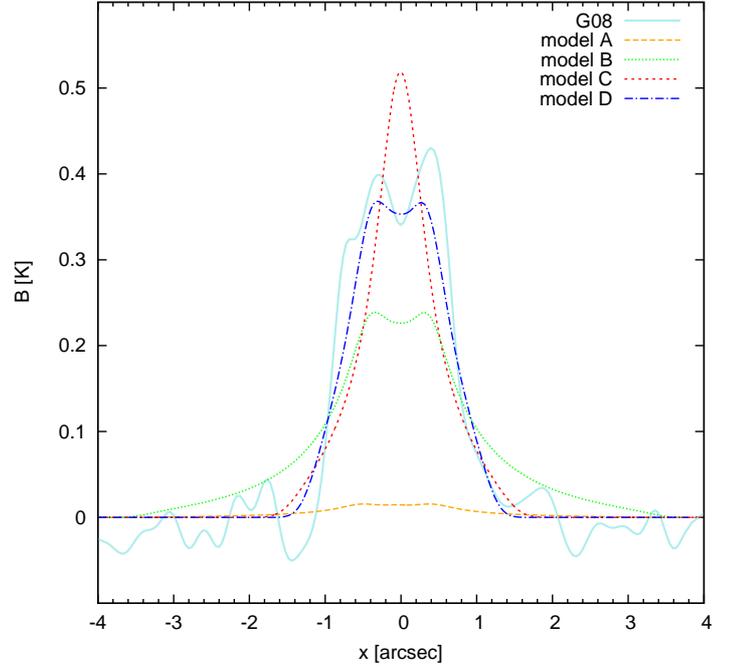}
  \caption{Direct comparison of horizontal cuts through the brightness map at $\lambda=1.3$mm of the deconvolved PdBI observation and our best-fit models.}
  \label{fig:mm_slices}
\end{figure}

The best-fit model C has a slightly steeper and colder radial temperature distribution than the previous model B with $q{\sim}0.43$ and $T_{100}{\sim}12$K for distances $R{\gtrsim}5$AU and therefore lies closer to the molecular line observations by P06, see Fig.~\ref{fig:best_temp}. Although its SED appears to be a good match to observation, the 1.3mm brightness map obtained from model C does not show a central dip, see Fig.~\ref{fig:mm_slices}. This finding suggests that it is improbable to simultaneously reproduce the SED and PdBI observations with a simple flared disk ansatz.

\subsection{Simultaneous fitting of SED and mm map}

The tradeoff model D is similar to parameter set B, apart from its smaller outer radius. The radial temperature distribution in the mid-plane is close to model B with $q{\sim}0.36$ and $T_{100}{\sim}19$K for $R{\gtrsim}5$AU. All deduced properties are comparable except for its mm brightness map, which features an improved shape and peak flux although the contrast of the double peak is still insufficient, but the differences are not significant. The optical depth of model D at $\lambda=1.3$mm is $\tau{\sim}1.8$ in the mid-plane and $\tau{\sim}0.9$ along the line of sight, so the disk is semitransparent in the mm-wavelength domain.

\subsection{Discussion}
Our study confirms previous results using an unbiased, extended disk model and a completely different 
minimization method, which is less susceptible to secondary minima. 
Nevertheless, none of the Markov chains of run B converged toward $\eta{\sim}1$, although W02 presented a well-matched 
simple flared disk model. The two domains with small and large $\eta$ appear to be separated by a 
"ridge" in the $\chi^2$-distribution that our specific implementation of the annealing schedule cannot overcome 
easily within the performed number of steps. 
To check the possibility of a simple flared disk solution, we fixed $\eta{=}1$ and found model C and similar parameter sets with 
$\chi^2_{\rm{SED}}$ comparable to model B. Their appearance at 1.3mm suggests that the deficit of mm emission toward 
the center of HH 30 is unlikely to be explained by an optical depth effect in a nearly edge-on simple flared disk, 
see Fig.~\ref{fig:mm_slices}. Therefore a change of mm opacity by a factor of ${\gtrsim}10$ at ${\sim}45$AU between 
an inner and outer domain is required, as discussed in G08. This result can be directly understood from the mm 
brightness image.

The typical dust temperature $T_{\rm{dust}}$ at 50AU in our best-fit models is ${\gtrsim}20$K, comparable to temperatures deduced from molecular line
observations (P06), see Fig.~\ref{fig:best_temp}. The surface brightness $T_{\rm{b}}$ of an optically thick emitter filling the whole beam is just its mean
temperature, therefore we can deduce from the actually observed peak brightness of $\hat{T}_{\rm{b}}\,{\lesssim}0.5$K that any optically thick emitter has to be
geometrically diluted by a factor of $T_{\rm{dust}}/\hat{T}_{\rm{b}}\,{\gtrsim}40$. The orientation of the beam resolves the disk of HH 30 radially and leaves
only vertical settling of dust as an explanation for an optically thick scenario. The dilution factor is accordingly given by the ratio of the projected disk thickness to the vertical resolution, $0.59"{\sim}83$AU at the distance of ${\sim}140$pc for Taurus. Thus, the \emph{apparent} disk thickness should be ${\sim}2$AU, incompatible with the derived inclination of ${\sim}83^\circ$, because $\sin 7^\circ{\sim}0.12$ and the diameter of the optically thick parts should be in the order of the distance between the peaks, i.e. ${\sim}100$AU. To still be able to explain the deficit of mm emission towards HH 30 by self-absorption, two effects have to be combined:\\

\begin{enumerate}
\item A very strong settling of larger grains to obtain a scale height $h{\sim}1$AU at $R{\sim}50$AU.
\item A substantial disk warp aligning the inner parts to appear nearly edge-on (${\gtrsim}88^\circ$), while
the outer disk, dominating the scattered light images, remains at an inclination of ${\sim}83^\circ$.
\end{enumerate}

While this scenario cannot be excluded, these peculiar features are beyond the scope of models tested so far, 
because all assume cylindrical symmetry. Given that inner holes have now been discovered in many disks, 
e.g. in LkCa 15 (Pi{\'e}tu et al. 2006), GM Aur (Dutrey et al. 2008; Hughes et al. 2009), and CB 26 (Sauter et al. 2009), a low-opacity cavity 
is a simple and viable explanation for the appearance of HH 30 at mm-wavelengths. Reanalysis by G08 of the $^{13}$CO data published in P06 shows that the lack of emission for ${\gtrsim}3$km/s can be explained by removing gas from the inner $(33{\pm}7)$AU. Anglada et al. (2007) proposed a wide binary to explain the jet wiggling that would clean the inner disk by tidal interaction as exemplified by G08. In this study we showed that a good fit of the SED and the double-peaked mm map can be achieved simultaneously if a depletion zone is included (Model D). Hence, this notion is now supported by several independent pieces of evidence.

Attempting to account for all known broadband measurements confirms a number of previously noted facts, and helps revealing their possible origins. Under our hypothesis of homogeneous dust properties, no single model can reproduce the ensemble of available data. A major problem is the outer radius of the dust disk. Resolved observations in the NIR and the SED are consistent with $R_{\rm{out}}{\sim}200..500$AU, while the continuum visibilities measured with the PdBI alone clearly indicate $R_{\rm{out}} = (135{\pm}15)$AU, a discrepancy that cannot be explained with a single dust species and suggests a radial dependency of the dust composition. A recently conducted high-resolution survey at mm-wavelengths with a sample of 10 TTauri disks found the same radial change in opacity (Guilloteau et al. 2011), indicating a negative gradient of mean grain size $\partial\left<a\right>/\partial R{<}0$ in the outer disk with populations of smaller particles present in this region. As pointed out by G08, the outer radius of the gas disk (${\sim}420$AU) is consistent with this interpretation. Further optimization should be continued in a follow-up study when additional high-resolution observations in the (sub)mm domain are available that reveal the spatial dependence of the spectral index $\beta_{\rm{mm}}$. Our results show that the mm observations described in G08 can be reproduced by a deficit of emission within ${\sim}45$AU, while the SED indicates in the MIR substantial amounts of warm dust in this region. We can explain the lack of flux from the inner zone at 1.3mm by an overall depletion of dust. However, the large grain size required to produce sufficient mm flux is inconsistent with chromaticity observed in scattered light images. Dust settling may explain this effect, because the scale height of large, and therefore mm emitting, grains is only loosely constrained. Another equally likely scenario is the migration of small dust grains coupled to the gas flow from the outer ring through the inner region, e.g. along streamers (Grinin et al. 2004; Kley et al. 2008; Tambovtseva et al. 2008), while larger grains decouple from this motion and are somehow effectively removed in the depletion zone. Either way, the inner region is not devoid of gas and dust as the SED in the MIR and the active jet clearly demonstrate. Only interferometers with sufficient sensitivity and sub-arcsecond
resolution like ALMA will be able to significantly reduce additional model degeneracies introduced by parameters describing an inhomogeneous grain size distribution, and disentangle whether vertical stratification is required in addition to radial segregation.

\section{Summary}

We conducted a parameter study to find an unbiased, self-consistent model of HH 30's disk able to predict all available continuum data. The investigation started with the $\alpha$-disk model and a standard dust composition as used in previous modeling efforts. Preliminary results exhibited incongruent fluxes and spectral indices in the MIR and for (sub)mm-wavelengths, leading to a refinement of the model. To compensate for the discrepancies between model predictions and observations, we added a variable grain size distribution and an attenuated region, thereby introducing larger grains and warm dust near the central source. The introduction of a depleted region and variable dust size added five parameters to our study and rendered an exhaustive search for the global best fit impractical, which led us to implement SA. Modification of the density by an attenuation $\eta$ was mainly selected for its simplicity and many configurations compatible with the observed flux are conceivable. Our model concentrates most of the dust near the mid-plane, causing higher optical depths in this region. Mid-infrared flux is hence mainly produced near the inner edge and in warm, upper layers because the inner parts toward the mid-plane are optically thick in this wavelength domain. Less massive configurations with lower optical depth may reproduce the observed flux in the MIR, therefore the deduced dust mass of the depleted region should be interpreted as an upper limit. 

Earlier work reported extensive ambiguities when fitting observations, even to the point that a single best-fit parameter set could not be given (W04), and we encountered the same obstacle, especially when fitting single data sets. 

The parameter set A found during optimization of NIR images reproduces the appearance of HH 30 in scattered light, but is not able to correctly predict any observation at longer wavelengths. Our inability to reproduce the contrast between secondary maximum and the obscuring belt in the NIR images may indicate the presence of additional features not present in our simple model, such as warps, an upwhirled dust cone, or asymmetrical illumination.

Degeneracies were also present during SED fitting, though new data sets have substantially improved the constraints on our predictions, most notably the observations by IRS/SST in the MIR. The parameter set of model B was found during optimization of the SED and replicates the flux in MIR quite well, but does not exhibit the correct appearance in the NIR and its mm brightness map is too extended, showing low peak brightness and little contrast. The overall agreement of observation and simulation at mm-wavelengths is only satisfactory for this model because the fluxes at 2.7mm and 3.4mm are underpredicted. We deduced confidence intervals for this model using Markov chains that sampled the vicinity of $\mathbf{a}_B$, see Table \ref{tab:best_fit}. 

To improve the model and reduce degeneracies we combined the optimization of two different classes of observations, photometry from MIR to mm-wavelengths and interferometric imaging at 1.3mm, and optimized with our new \emph{staggered} MCMC method that does not weigh individual $\chi^2_\alpha$ during optimization. This approach did remove some degeneracies and the resulting best-fit model D has an improved mm brightness map without compromising the SED. 

The findings of our parameter study extend the results of previous works, see Tables \ref{tab:previous}, \ref{tab:modelA_prop}, and \ref{tab:best_fit} for comparison. This is not self-evident because our modeling effort was not biased by fixing parameters \emph{a priori} and the standard ansatz was modified by adding an attenuated inner zone, which introduced new degeneracies.

A greater refinement of the model appears to be necessary to satisfactorily explain all observational data at once. The most prominent discrepancies between model A and the other best-fit models lie in the grain size distribution and hence dust mass and chromaticity. This suggests that a disk of larger settled grains embedded in an overlaying disk of smaller dust grains may be able to reconcile theory and observation. A model accounting for spatially inhomogeneous grain size distributions has to include at least two different dust species and several new parameters, e.g. D'Alessio et al. (2006), Sauter \& Wolf (2010). Unfortunately, at the moment we lack observational data to disentangle the additional degeneracy, which significantly reduces the plausibility of any such model. Nevertheless, our results are consistent with an inner depletion zone of $(45{\pm}5)$AU radius, slightly larger than the hole proposed in G08, and suggest dust growth, vertical settling, and radial segregation as mm opacity changes significantly at ${\sim}140$AU. 
 
High-resolution imaging of HH 30 in different wavelength domains with next-generation observatories, such as the interferometer ALMA and the space-born JWST, will break the inherent model degeneracy and help to uncover the structure of the inner region, map the spatial dependency of dust properties, and refine our understanding of this fascinating object and its evolutionary stage in the coming years.

\begin{acknowledgements}
Images in the NIR used for this study were made with the NASA/ESA \emph{Hubble Space Telescope}, and obtained from the Hubble Legacy Archive, which is a collaboration between the Space Telescope Science Institute (STScI/NASA), the Space Telescope European Coordinating Facility (ST-ECF/ESA) and the Canadian Astronomy Data Centre (CADC/NRC/CSA).

The SED was observed in the MIR with the \emph{Spitzer Space Telescope}, and obtained from the NASA/IPAC Infrared Science Archive, both of which are operated by the Jet Propulsion Laboratory, California Institute of Technology under contract with the National Aeronautics and Space Administration.

We would like to thank E. Furlan for the re-reduced IRS dataset and her comments on the reduction process, and F. Madlener for patiently correcting the manuscript.

\end{acknowledgements}

\end{document}